\documentclass[submission, PhysProc]{SciPost}

\hypersetup{
    colorlinks,
    linkcolor={red!50!black},
    citecolor={blue!50!black},
    urlcolor={blue!80!black}
}

\usepackage[bitstream-charter]{mathdesign}
\DeclareSymbolFont{usualmathcal}{OMS}{cmsy}{m}{n}
\DeclareSymbolFontAlphabet{\mathcal}{usualmathcal}

\def\dd{{\,\rm d}}
\def\bis{{\prime\prime}}
\newcommand{\half}{{\textstyle\frac{1}{2}}}
\newcommand{\threehalf}{{\textstyle\frac{3}{2}}}

\def\vecb{{\boldsymbol b}}
\def\veck{{\boldsymbol k}}
\def\veckappa{{\boldsymbol \kappa}}
\def\vecq{{\boldsymbol q}}
\def\vecr{{\boldsymbol r}}
\def\Order{{\cal O}}
\renewcommand{\Re}{{\thinspace\rm Re\thinspace}}
\renewcommand{\Im}{{\thinspace\rm Im\thinspace}}

\begin{document}

\begin{center}{\Large \textbf{
Roy Glauber and Asymptotic Diffraction Theory\\
}}\end{center}

% TODO: write the author list here. Use first name (+ other initials) + surname format.
% Separate subsequent authors by a comma, omit comma and use "and" for the last author.
% Mark the corresponding author with a superscript star.
\begin{center}
Per Osland\textsuperscript{1$\star$}
\end{center}

\begin{center}
{\bf 1} Department of Physics and Technology, University of Bergen, \\
Postboks 7803, N-5020  Bergen, Norway
\\
${}^\star$ {\small \sf Per.Osland@uib.no}
\end{center}

\begin{center}
\today
\end{center}

% For convenience during refereeing (optional),
% you can turn on line numbers by uncommenting the next line:
%\linenumbers

\definecolor{palegray}{gray}{0.95}
\begin{center}
\colorbox{palegray}{
  \begin{tabular}{rr}
  \begin{minipage}{0.05\textwidth}
    \includegraphics[width=14mm]{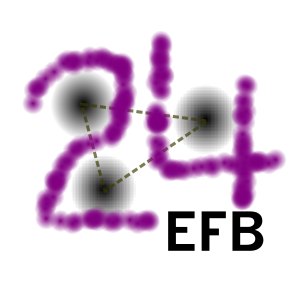}
  \end{minipage}
  &
  \begin{minipage}{0.82\textwidth}
    \begin{center}
    {\it Proceedings for the 24th edition of European Few Body Conference,}\\
    {\it Surrey, UK, 2-4 September 2019} \\
    %\doi{10.21468/SciPostPhysProc.2}\\
    \end{center}
  \end{minipage}
\end{tabular}
}
\end{center}

\section*{Abstract}
{\bf
% TODO: write your abstract here.
This is a review of Glauber's asymptotic diffraction theory, in which diffractive scattering is described in terms of interference between semiclassical amplitudes, resulting from a stationary-phase approximation. Typically two such amplitudes are sufficient to accurately describe elastic scattering, but the stationary points are located at complex values of the impact parameter. Their separation controls the interference pattern, and their offsets from the real axis determine the overall fall-off with momentum transfer. Asymptotically, at large momentum transfers, the stationary points move towards singularities of the profile function. I also include some reminiscences from our collaboration.
}

\vspace{10pt}
\noindent\rule{\textwidth}{1pt}
\tableofcontents\thispagestyle{fancy}
\noindent\rule{\textwidth}{1pt}
\vspace{10pt}

\section{Introduction}
\label{sec:intro}
% TODO: write your article here.
I had the pleasure of working with Professor Roy Glauber on and off for more than 40 years, starting as a postdoctoral fellow at Harvard in 1976, and up to the recent completion of our book \cite{Diffraction}.
In this note, I first give some brief comments to his vita, then outline representative elements of our book, before concluding with some reminiscences from our many years of interactions.

%%%%%%%%%%%%%%%%%%%%%%%%%%%%%%%%%%%%%%%%%%%%%%%%%%
\section{A Personal View of Roy Glauber's Vita}
\begin{figure}[h]
\begin{center}
\includegraphics[scale=0.60]{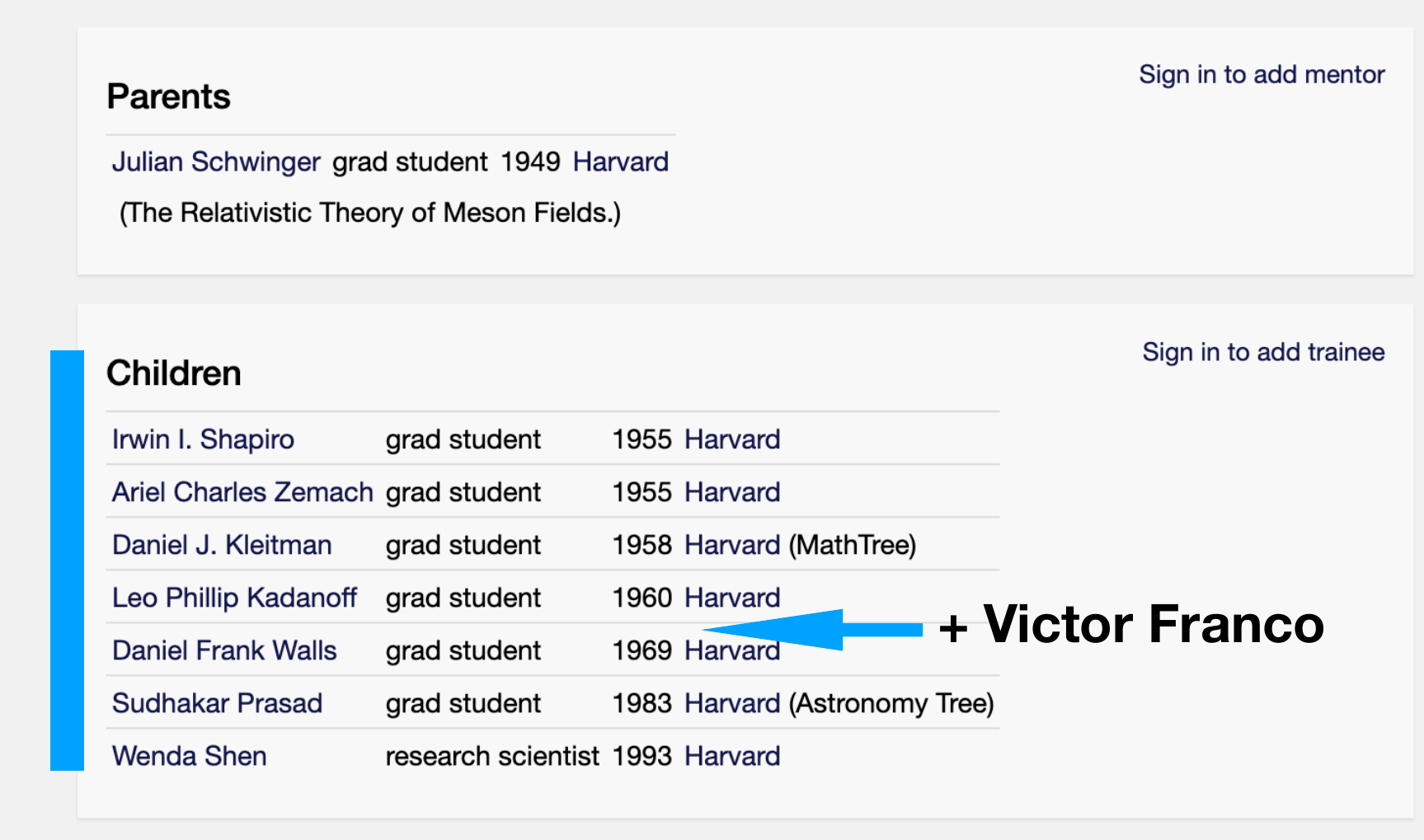}
\end{center}
\caption{Edited excerpt from ``Academic tree'' webpage on Roy Glauber \cite{Academictree}.}
\label{children}
\end{figure}

%%%%%%%%%%%%%%%%%%%%%%%%%%%%%%%%%%%%%%%%%%%%%%%%%%
In the preparation of this talk, I consulted the ``Academic tree'' webpage 
\cite{Academictree}.
Here, I found interesting information on the academic ancestors of his PhD supervisor, Julian Schwinger, as well as a list of ``children'', basically PhD students. However, the latter were mostly related to his activity in Quantum Optics, for which he was awarded one half of the 2005 Nobel Prize in Physics, with the citation: ``for his contribution to the quantum theory of optical coherence''.
Missing from that list was a student who was very much involved in Roy's early work on multiple scattering theory, namely Victor Franco (see Fig.~\ref{children}). However, the list reminds us of Roy's impact also on other fields (other than Quantum Optics and Scattering Theory), like Mathematics and Statistical Physics.

%%%%%%%%%%%%%%%%%%%%%%%%%%%%%%%%%%%%%%%%%%%%%%%%%%
%%%%%%%%%%%%%%%%%%%%%%%%%%%%%%%%%%%%%%%%%%%%%%%%%%

%%%%%%%%%%%%%%%%%%%%%%%%%%%%%%%%%%%%%%%%%%%%%%%%%%
\begin{figure}
\begin{center}
\includegraphics[scale=0.7]{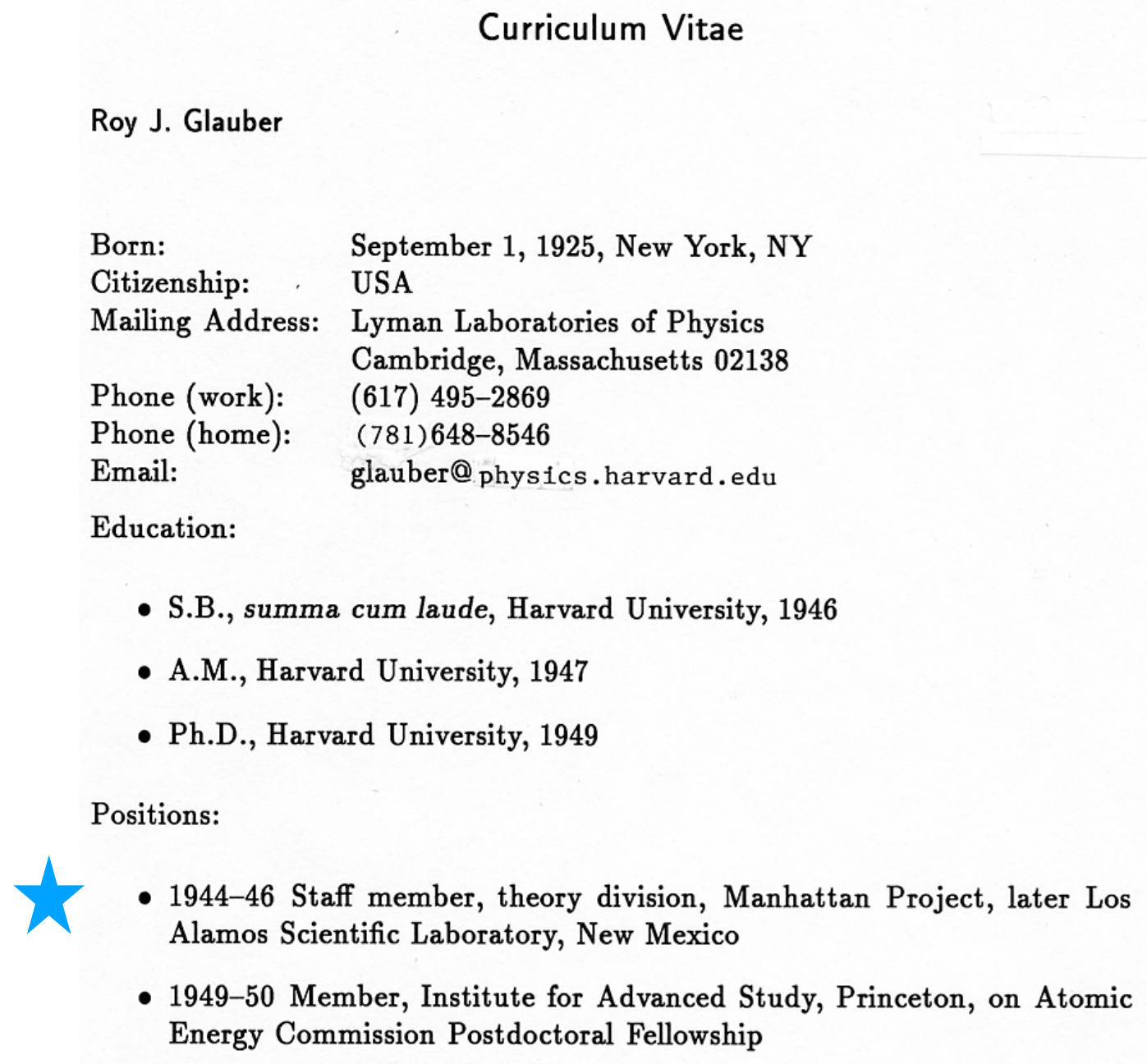}
\end{center}
\caption{From Roy Glauber's CV dated '97, with his Manhattan Project position highlighted.}
\label{roy-cv-1}
\end{figure}
%%%%%%%%%%%%%%%%%%%%%%%%%%%%%%%%%%%%%%%%%%%%%%%%%%

I was not aware of Roy's participation in the Manhattan Project until I got hold of a copy of his CV in 2007 for the purpose of nominating him for an honorary degree. He had never mentioned it, so I was left wondering whether he perhaps had mixed feelings about it. Anyway, it is listed as his first appointment, 18 years old, in the 1997-version of his CV, see Fig.~\ref{roy-cv-1} (and described in some detail in his biography at the Nobel Foundation website \cite{GlauberBib}). After the war, he returned to Harvard, completed his PhD with Schwinger in 1949, and after a few years at Princeton (1949--1951) and Caltech (1951--1952), he returned to Harvard in 1952. Starting as a lecturer, he was subsequently promoted to assistant and associate professor, and then full professor in 1962 and Mallinckrodt Professor of Physics in 1976.

The first papers listed in his CV are from his contributions to the Manhattan Project: {\it The Critical Masses of Tamped Spheres} (1944--1945), {\it The Stopping of Multiplication in Expanding, Tamped Spheres} (1944--1945),  and {\it Neutron Diffusion in Spherical Media of Radially Varying Density} (1944--1945). His PhD thesis (1949) was on {\it The Relativistic Theory of Meson Fields}, and then in the early 50s he started writing papers on electron diffraction and scattering theory. 

The famous ``multiple-scattering'' papers {\it Deuteron Stripping Processes at High Energies} and {\it Cross Sections on Deuterium at High Energies} appeared in 1955, and it is natural to speculate whether the multiple-scattering branch of his work grew out of the experience from the Manhattan project. Then, in 1959, came his Boulder Lectures: {\it High Energy Collision Theory}, which for many became a ``bible'' of hadron-nucleus scattering.

The work on Quantum Optics, for which he was awarded the Nobel Prize, started in 1962, with the publication of {\it Photon Correlations} \cite{Glauber1963} and follow-up papers. Gradually, this became his main interest and activity. But he did not abandon the field of scattering theory, something I was to benefit from.

%%%%%%%%%%%%%%%%%%%%%%%%%%%%%%%%%%%%%%%%%%%%%%%%%%
\section{Asymptotic Diffraction Theory}
\label{sect:diffraction}
%%%%%%%%%%%%%%%%%%%%%%%%%%%%%%%%%%%%%%%%%%%%%%%%%%

\begin{figure}
\begin{center}
\vspace*{-10mm}
\includegraphics[scale=0.5]{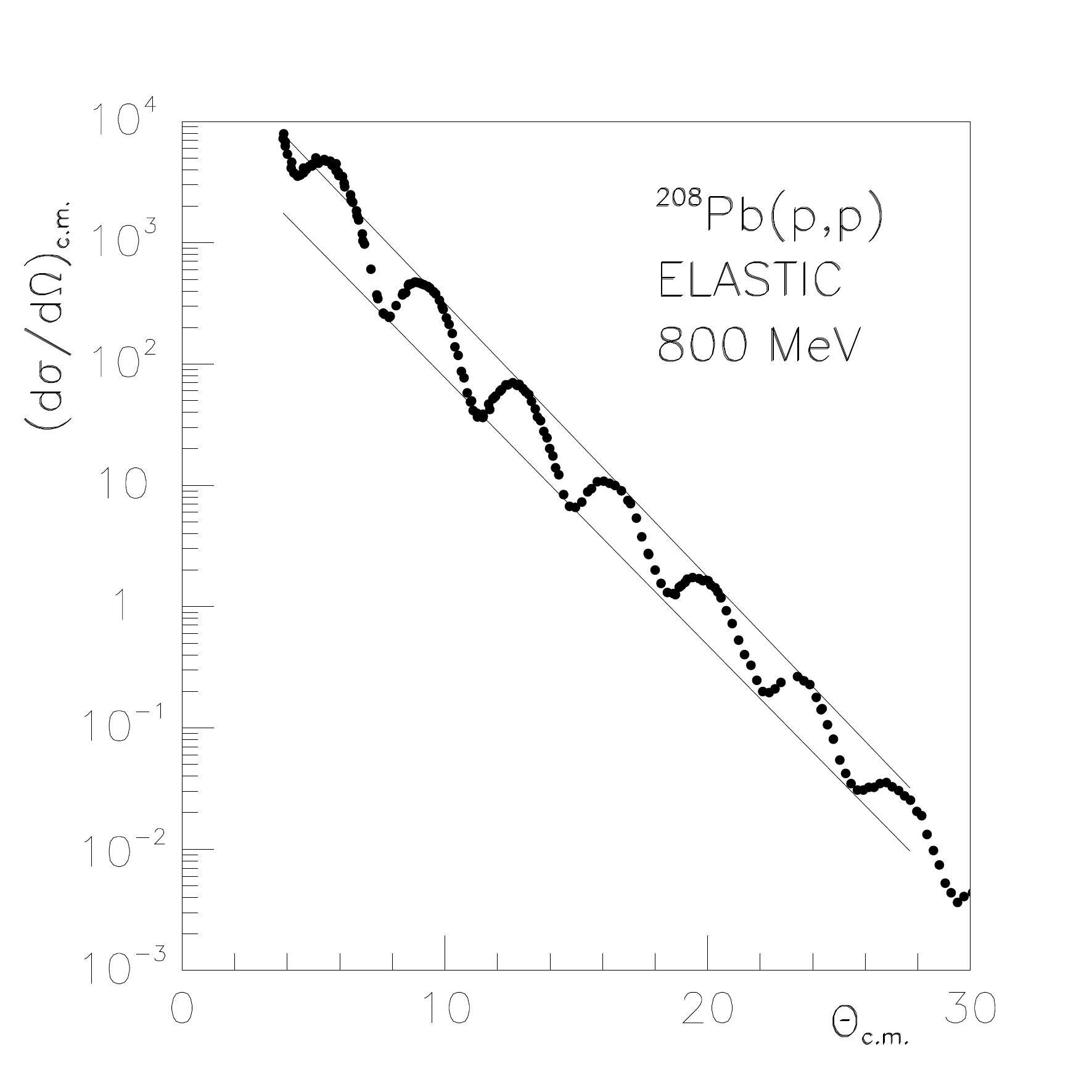}
\end{center}
\vspace*{-10mm}
\caption{
Differential cross section for elastic proton-lead scattering at 800~MeV
\cite{Hoffmann:1980kg}.}
\label{Fig:int:pb-el}
\end{figure}

Around 1980, various proton elastic and inelastic scattering experiments were performed at the Los Alamos National Laboratory, aimed at the investigation of nuclear structure. For a heavy target, like ${}^{208}\text{Pb}$, the differential cross section was rather monotonous, roughly given by a periodic oscillatory pattern that had an exponentially falling envelope, as illustrated in Fig.~\ref{Fig:int:pb-el}.
Returning to Harvard in 1981, I learned that Roy and Marek Bleszynski had been working on a simplified description of proton-nucleus scattering, an activity I immediately joined. The work was relevant to those experiments at  Los Alamos, in particular, to the activity of G.~Hoffmann and collaborators \cite{Hoffmann:1980kg}.\footnote{The work may have been inspired by earlier work using asymptotic approximations to the Bessel function \cite{Amado:1979st}.}
Roy was to spend the summer of that year at CERN, and before his leaving Cambridge, we quickly patched together a manuscript on our ``simplified'' understanding of these differential cross sections \cite{Bleszynski:1981fk}, to be hand delivered by Roy to the Physics Letters~B editor at CERN.

%%%%%%%%%%%%%%%%%%%%%%%%%%%%%%%%%%%%%%%%%%%%%%%%%%
%%%%%%%%%%%%%%%%%%%%%%%%%%%%%%%%%%%%%%%%%%%%%%%%%%

The study was further developed during another stay at Harvard in the mid-80s, and during visits by Roy to Bergen in the early 90s. It then lay dormant until 2006, at which point also a conversion of all illustrations to electronic versions was started.

The theory is based on the Kirchhoff integral representation of Fraunhofer diffraction \cite{Boulder},
\begin{equation} \label{Eq:diff-ampl}
f(\veck',\veck)
=\frac{ik}{2\pi}\int 
e^{-i\vecq\cdot\vecb}\bigl\{1-e^{i\chi(\vecb)}\bigr\}\dd^2b.
\end{equation}
Our sign convention\footnote{In Ref.~\cite{Bleszynski:1981fk}. the opposite sign convention for $\vecq$ was used.} is such that $\vecq=\veck^\prime-\veck$.
Here, the ``1'' only contributes a delta function in the forward direction, and our approach was to evaluate the remaining integral by the method of stationary phase. The points of stationary phase are given by
\begin{equation} \label{Eq:as-diff-stat-phase}
\vecq=\nabla_b\chi(\vecb).
\end{equation}
While $\vecq$ is a real quantity, the phase shift function is in general complex. Thus, the solutions for $\vecb$ will in general be located off the real axes in the two-dimensional $\vecb$ plane.

If $\vecb_0$ is a point of stationary phase, the stationary-phase approximation to the integral (\ref{Eq:diff-ampl}) is obtained by expanding $\chi(\vecb)$ to second order in the deviations 
\begin{equation}
\Delta\vecb=\vecb-\vecb_0=(\Delta b_x,\Delta b_y)=(x,y),
\end{equation}
and performing the two Gaussian integrals over $x$ and $y$. Here, $x$ is along the positive $q$ direction and $y$ is along the normal to the scattering plane. One thus finds for the amplitude corresponding to one particular point of stationary phase, $\vecb_0$  \cite{Diffraction}
\begin{equation} \label{Eq:as-diff-f1}
f_0(\veck^\prime,\veck)
=\frac{k}{i}\left(\frac{-b_{0\,x}}{q\chi^\bis(b_{0\,x})}\right)^\half
e^{-iqb_{0\,x}+i\chi(b_{0\,x})},
\end{equation}
where azimuthal symmetry has been assumed and the second derivative in $y$ has been expressed in terms of $q$ and the $x$-component of $\vecb_0$. 
Thus, it is given by the integrand at the point of stationary phase, multiplied by a ``prefactor'' determined by the second derivatives.
Typically, there will be contributions from two or more points of stationary phase.

\subsection{Classical scattering}
Before illustrating applications of the approximation (\ref{Eq:as-diff-f1}) to some representative phase shift functions, let us first consider a particle in a classical potential $V(\vecr)$, where $\vecr=\hat\veckappa z +\vecb$, with $\veckappa=\half(\veck+\veck^\prime)$. It will experience a transverse force $-\nabla_b V(\vecb+\hat\veckappa z)$. Integrated over time ($\dd t=\dd z/v$), we get the
transfer of momentum to the scattered particle,
\begin{equation} \label{Eq:as-diff-stat-cond}
\hbar(\veck^\prime-\veck)=\hbar\vecq=\hbar\nabla_b\chi(\vecb),
\end{equation}
with $\chi$ related to the potential by
\begin{equation} \label{Eq:diff-chi-def}
\chi(\vecb)=-\frac{1}{\hbar v}
\int_{-\infty}^\infty V(\vecb+\hat\veckappa z)\dd z.
\end{equation}
We note that the classical condition (\ref{Eq:as-diff-stat-cond}) is just the condition of stationary phase, Eq.~(\ref{Eq:as-diff-stat-phase}). It is easy to see \cite{Diffraction} that the classical cross section will be given by the square of the modulus of the expression (\ref{Eq:as-diff-f1}).

\subsection{Example~1. Coulomb scattering}
\label{subsect:Coulomb-point}
Consider scattering from a point Coulomb charge,
\begin{equation} \label{Eq:exa:Coulomb-pot}
V(\vecr)=\frac{Ze^2}{4\pi r}.
\end{equation}
The phase shift function is actually infinite, but if we introduce a screening radius $R$ (for simplicity, we consider a sharp cut-off), it can be written as
\begin{align}
\chi(b)&=-\frac{2Ze^2}{4\pi\hbar v}
\int_b^R\frac{\dd r}{\sqrt{r^2-b^2}} \\
&=2\eta\log\left(\frac{b}{2R}\right)+\Order\left(\frac{b^2}{R^2}\right),
\end{align}
where $v$ is the projectile velocity, and we have introduced the Sommerfeld parameter \cite{Sommerfeld}
\begin{equation} \label{Eq:simple-Sommerfeld}
\eta=\frac{Ze^2}{4\pi\hbar v}.
\end{equation}
There is only one stationary point, given by
\begin{equation}
b_x=\frac{2\eta}{q}.
\end{equation}
It is on the real $b_x$-axis, and has the same sign as $\eta$, i.e., positive for repulsive forces and negative
for attractive ones. The resulting amplitude is
\begin{equation} \label{Eq:ex:Coulomb-ampl}
f(\veck^\prime,\veck)=\frac{2\eta k}{q^2}
\exp\biggl\{-2i\eta\log\frac{qR}{\eta}-2i\eta+\frac{i\pi}{2}\biggr\},
\end{equation}
and the differential cross section, $\dd\sigma/\dd\Omega=|f(\veck^\prime,\veck)|^2$, is {\it identical} to the Rutherford cross section (the phase is unobservable).

We shall in section~\ref{sect:Coulomb-distributed} consider scattering from a distributed charge, leading to interference phenomena and also a rainbow.

\subsection{Example~2. Paired trajectories, rainbows}
\begin{figure}
\begin{center}
\vspace*{-10mm}
\includegraphics[scale=0.45]{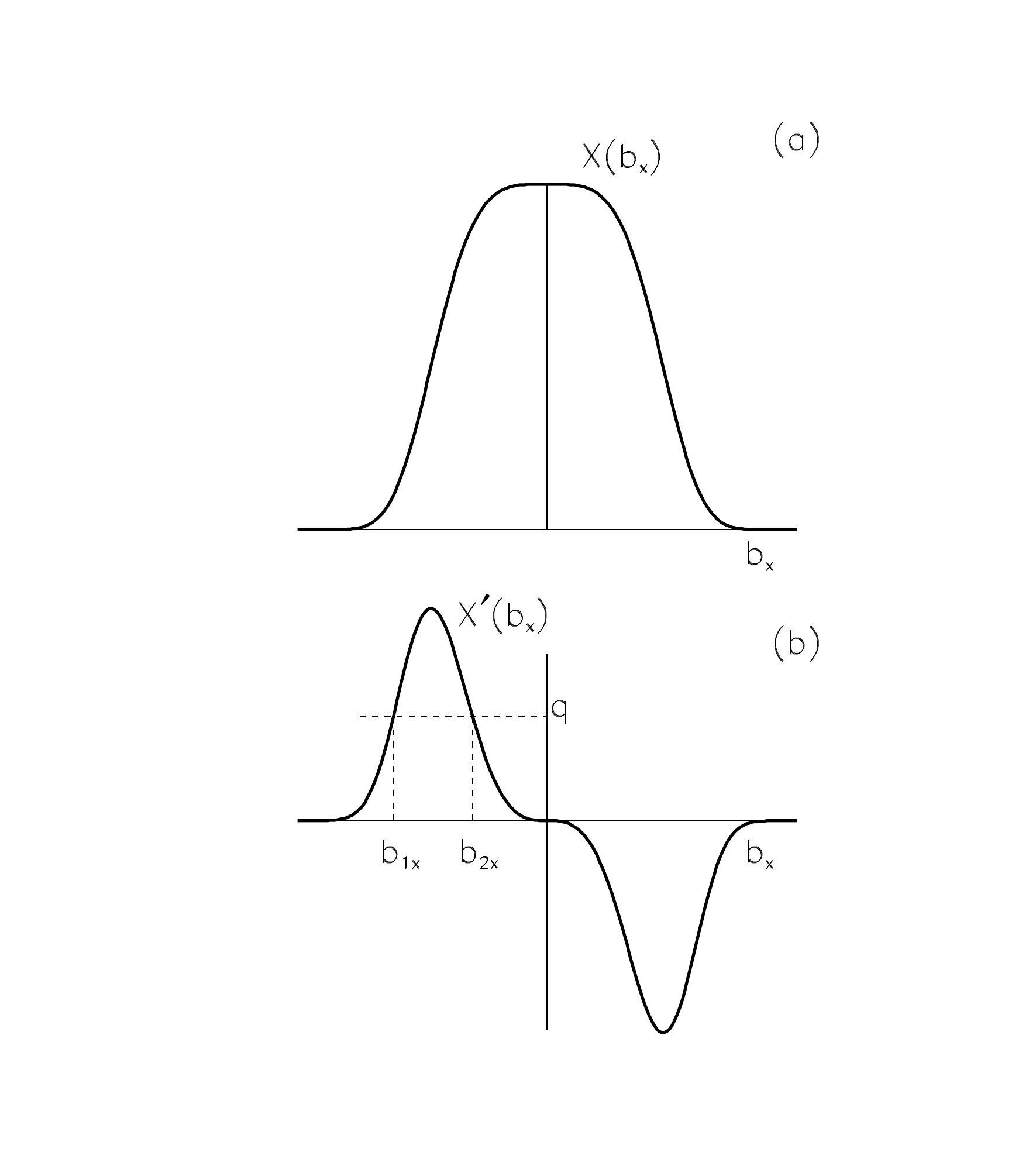}
\end{center}
\vspace*{-10mm}
\caption{
(a) A simple, real, phase shift function $X(b_x)$ and (b) its derivative $X^\prime(b_x)$. For values of $q$ that are not too large, there are paired trajectories at $b_{1x}$ and $b_{2x}$. The stationary points are located at negative values of $b_x$, since a positive phase shift function corresponds to attraction.}
\label{Fig:paired-trajectories}
\end{figure}

Let us consider a simple, real, phase shift function, which for $b_y=0$ has a shape given by $X(b_x)=\chi(b_x,0)$, qualitatively as illustrated in Fig.~\ref{Fig:paired-trajectories}. It is even in $b_x$, and its derivative will thus be an odd function. In this example, it is seen that for $q$ less than some critical value, there will be two solutions to the stationary phase condition (\ref{Eq:as-diff-stat-phase}). There will be two interfering amplitudes, each of the form (\ref{Eq:as-diff-f1}). 
%%%%%%%%%%%%%%%%%%%%%%%%%%%%%%%%%%%%%%%%%%%%%%%%%%
%%%%%%%%%%%%%%%%%%%%%%%%%%%%%%%%%%%%%%%%%%%%%%%%%%

However, for $q$ equal to some critical value (denoted $q_R$), the two stationary points merge, and the amplitude, inversely proportional to the square root of the second derivative, diverges. This is a rainbow. For $q>q_R$, there is apparently no solution. This is however no longer true if we allow for complex values of the impact parameter $b_x$. The two roots that for $q<q_R$ were located on the real axis, will for $q>q_R$ move into the complex $b_x$-plane, one above and one below the real axis.\footnote{Mathematical note: we assume that $X^\bis(b_x)$ is analytic and has a simple zero at this point.} It is easy to see \cite{Diffraction} that (with our sign convention for the momentum transfer) it is only the root $b_x$ that is {\it below} the real axis that will be encountered along the path of integration over $b_x$.

Beyond the rainbow point ($q>q_R$), since the stationary point moves into the complex $b_x$ plane, the amplitude gets a factor $\exp(\Im b_xq)$ and the differential cross section will fall off exponentially. This is no surprise, since we are in a region that is classically forbidden.

One can easily imagine phase shift functions whose derivatives have more extrema, and thus more rainbows \cite{Diffraction}.

\subsection{Example~3. Absorption}
Inelastic scattering, or absorption, can be represented by a complex phase shift function. As a simple illustration of the associated elastic scattering, let $\chi(b)$ be pure imaginary, and have a shape similar to that shown in Fig.~\ref{Fig:paired-trajectories}. Its derivative, on the real axis, will again be like in Fig.~\ref{Fig:paired-trajectories}. However, it will be pure imaginary, and thus nowhere satisfy Eq.~(\ref{Eq:as-diff-stat-phase}). If we move off the real axis, the situation changes dramatically.

For illustration, consider the inverse $\cosh$ phase shift function,
\begin{equation}
\chi(b)=\frac{\chi_0}{\cosh(b/\beta)},
\end{equation}
with $\chi_0$ purely imaginary.
We need the derivative of this function, with respect to $b_x$, for $b_y=0$. Rather than writing this out, let us just note that for $b_y=0$ we have
\begin{equation}
X(b_x)=\frac{2\chi_0}{e^{b_x/\beta}+e^{-b_x/\beta}}.
\end{equation}
This has simple poles for 
\begin{equation}
\frac{b_x}{\beta}=\half\log(-1)=\frac{\pm i}{2}\{\pi,\, 3\pi,\text{etc.}\}.
\end{equation}
Consequently, the first derivative will have double poles at these locations, and at large $q$, the stationary points will approach the poles. Actually, since the poles are on the imaginary axis, it is only the one nearest to the real axis, $b_x^\text{pole}=-i\pi\beta/2$, that is relevant for the path of integration.

With $X(b_x)$ having a simple pole at $b_x^\text{pole}$, the derivative will have a double pole. With 
\begin{equation}
b_x=b_x^\text{pole}+\rho e^{i\phi}, 
\end{equation}
in the neighbourhood of this pole we have
\begin{equation}
X^\prime(b_x)\propto\frac{\chi_0\sinh(b_x/\beta)}{\rho^2 e^{2i\phi}},
\end{equation}
with $\sinh(b_x/\beta)$ predominantly imaginary.
For the whole expression to be real, and satisfy Eq.~(\ref{Eq:as-diff-stat-phase}), we need $2\phi=n\pi$ (since $\chi_0$ is pure imaginary). This example will thus at large momentum transfers give two points of stationary phase, located in the lower complex plane, at an angle of $\pm\pi/2$ with respect to the positive imaginary axis. (The requirement that $X^\prime(b_x)$ be real and {\it positive} removes solutions with an odd multiple of $\pi$.)
Let us denote them $b_{x\alpha}$ and $b_{x\beta}$.

The facts that 
\begin{enumerate}
\item
there are two points of stationary phase, and
\item
these are located below the real axis,
\end{enumerate}
mean that (1) the differential cross section will exhibit periodic oscillations, with a period determined by the separation of the two stationary points along the real axis, $\Re(b_{x\alpha}-b_{x\beta})$, and (2) the envelope of the differential cross section will be exponentially damped, with a slope given by $\Im b_{x\alpha}=\Im b_{x\beta}$. The two points of stationary phase will in this case be symmetrically located with respect to the imaginary axis, and the amplitudes will be of equal magnitude.

If $\chi_0$ were not pure imaginary, the two points of stationary phase would have been located at different offsets from the real axis, $\Im b_{x\alpha}\neq\Im b_{x\beta}$. The two amplitudes $f_\alpha(\veck^\prime,\veck)$ and $f_\beta(\veck^\prime,\veck)$ would then have different exponential fall-offs with $q$, one amplitude would eventually dominate, and the oscillations would not persist at high values of $q$.

\subsection{Example~4. Coulomb scattering from a distributed charge}
\label{sect:Coulomb-distributed}
For an extended charge distribution, there will be at least two points of stationary phase, and thus interference phenomena. We shall illustrate this for a Gaussian distribution, modified by a polynomial prefactor,\footnote{The coefficient $\alpha$ should not be confused with the point of stationary phase labelled $\alpha$.}
\begin{equation} \label{Eq-coul-gauss-charge}
\rho(r)=\frac{1}{\pi^{3/2}\beta^3}\frac{1}{1+\threehalf\alpha}
\biggl(1+\alpha\frac{r^2}{\beta^2}\biggr)e^{-r^2/\beta^2},
\end{equation}
as might be encountered in shell-model descriptions of light nuclei.
The phase shift function will involve an exponential integral, but its derivative (for $b_y=0$) is rather simple,
\begin{equation}
X^\prime(b_x)=\frac{2\eta}{b_x}
\biggl[1-\biggl(1+\frac{\alpha}{1+\threehalf\alpha}\,
\frac{b_x^2}{\beta^2}\biggr)
e^{-b_x^2/\beta^2}\biggr].
\end{equation}
In addition to a ``far-out'' stationary point, $b_x\sim 2\eta/q$ (cf.\ section~\ref{subsect:Coulomb-point}), there is now also a stationary point at the scale defining the charge distribution, $b_x=\Order(\beta)$. The corresponding two amplitudes will create an interference pattern. Furthermore, there is a point $b_{x,R}$ where $X^\bis(b_{x,R})=0$, i.e., a rainbow point. The corresponding momentum transfer will be $q_R=c\,\eta/\beta$, with $c$ a function of the coefficient $\alpha$ in Eq.~(\ref{Eq-coul-gauss-charge}).

%%%%%%%%%%%%%%%%%%%%%%%%%%%%%%%%%%%%%%%%%%%%%%%%%%
\begin{figure}
\begin{center}
\includegraphics[scale=0.5]{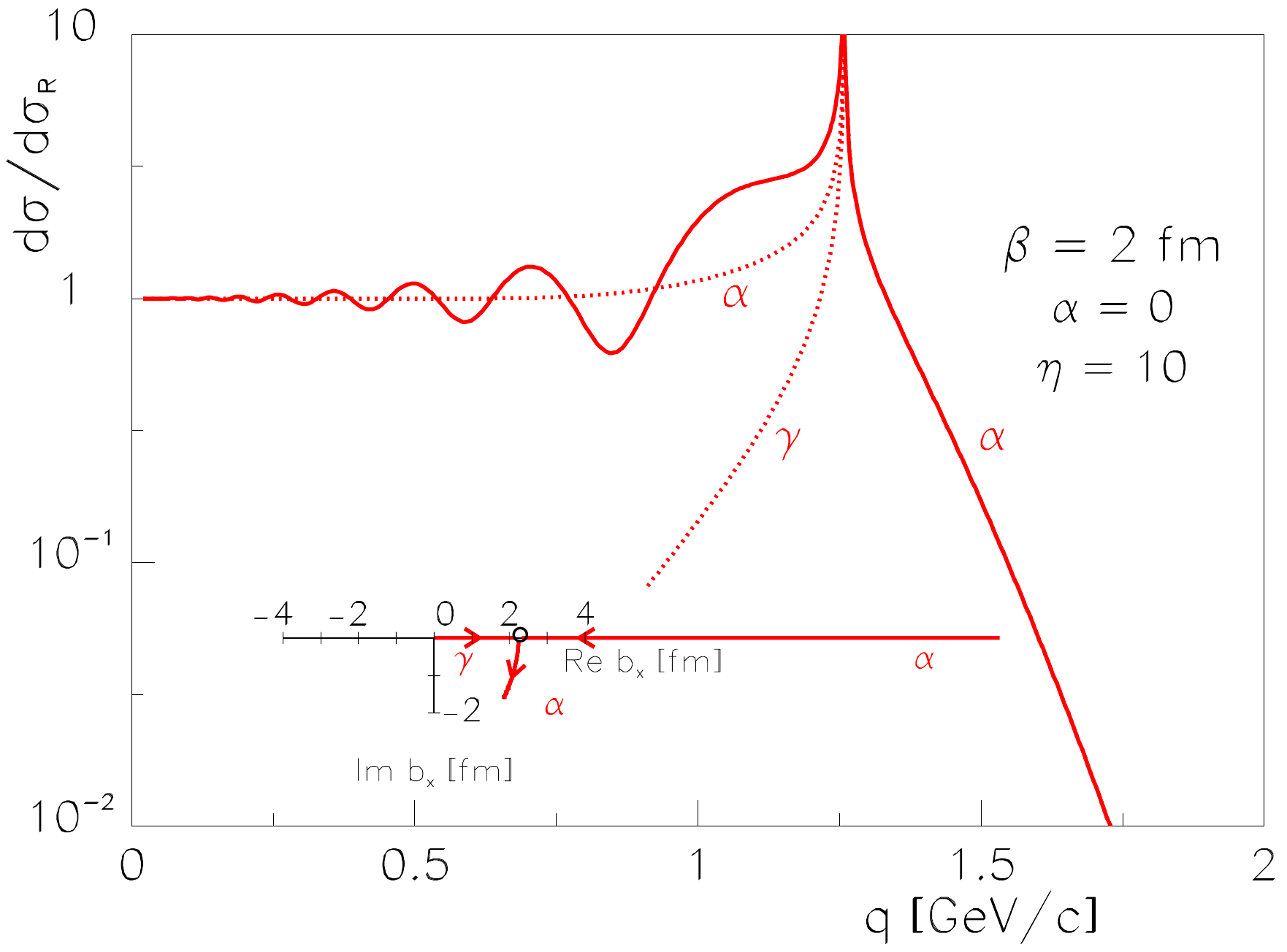}
\end{center}
\vspace*{-5mm}
\caption{
Scattering from the charge distribution (\ref{Eq-coul-gauss-charge}), normalized to the Rutherford cross section. The dotted curves show the individual contributions from the two stationary points. The inset shows how the stationary points move with increasing momentum transfer.}
\label{Fig:Coul-distribution}
\end{figure}
%%%%%%%%%%%%%%%%%%%%%%%%%%%%%%%%%%%%%%%%%%%%%%%%%%

In Fig.~\ref{Fig:Coul-distribution} we show the resulting differential cross section for a rather large value of the Sommerfeld parameter, $\eta=10$, normalized to the Rutherford cross section. Also shown, are the partial contributions of the two stationary points, labelled $\alpha$ and $\gamma$. The cross section exhibits a rainbow singularity, located at (in this case, for $\alpha=0$) $q_R=1.276\,\eta/\beta$. The inset shows the locations of the two stationary points, labelled $\alpha$ and $\gamma$, for a range of the momentum transfers, with the arrows indicating how they move with increasing values of momentum transfer. Beyond the rainbow point, the point labelled $\alpha$ moves down into the complex $b_x$-plane, giving an exponentially damped contribution, whereas the other, labelled $\gamma$, moves up into the complex plane, evading the path of integration.

Two aspects of the interference pattern are worth noting. (i) The period of oscillation increases as $q$ increases towards $q_R$. This is caused by the fact that the two points of stationary phase approach each other. (ii) The amplitude of the oscillation increases towards the rainbow point. This is obviously a reflection of the fact that the two amplitudes become more comparable in magnitude, as also seen from the two dotted curves in Fig.~\ref{Fig:Coul-distribution}.

%%%%%%%%%%%%%%%%%%%%%%%%%%%%%%%%%%%%%%%%%%%%%%%%%%
\section{Conclusion}
%%%%%%%%%%%%%%%%%%%%%%%%%%%%%%%%%%%%%%%%%%%%%%%%%%
The asymptotic evaluation of the diffraction integral (\ref{Eq:diff-ampl}) provides a rather precise approximation to the numerically ``exact'' result (see Ref.~\cite{Diffraction}). The contribution from integrating over the whole scattering plane ($\int\dd^2b$) can be represented by contributions from a small number of stationary phase points. The number and location of these points of stationary phase offers valuable qualitative insight into the scattering process. At large momentum transfers, the points of stationary phase will approach the singularities (normally, in the complex plane) of the phase shift function.

\section*{Acknowledgements}
It is a pleasure to thank the organizers, in particular Professors Natalia Timofeyuk and Jim Al-Khalili, for setting up a very interesting conference, and Professor Colin Wilkin for his initiative towards organizing a special session dedicated to the memory of Roy Glauber and his seminal contributions to this field. 
Furthermore, I would like to thank Marek Bleszynski and Lanny Ray for correspondence.
The figures \ref{Fig:int:pb-el}, \ref{Fig:paired-trajectories} and \ref{Fig:Coul-distribution} are reprinted from Ref.~\cite{Diffraction} with kind permission from Cambridge University Press.

% TODO: include author contributions

% TODO: include funding information
\paragraph{Funding information}
This work is supported in part by the Research Council of Norway, \\
http://dx.doi.org/10.13039/501100005416, contract number 255182.

\begin{appendix}

%%%%%%%%%%%%%%%%%%%%%%%%%%%%%%%%%%%%%%%%%%%%%%%%%%
\section{Personal reminiscences}
\begin{figure}
\begin{center}
\includegraphics[scale=0.6]{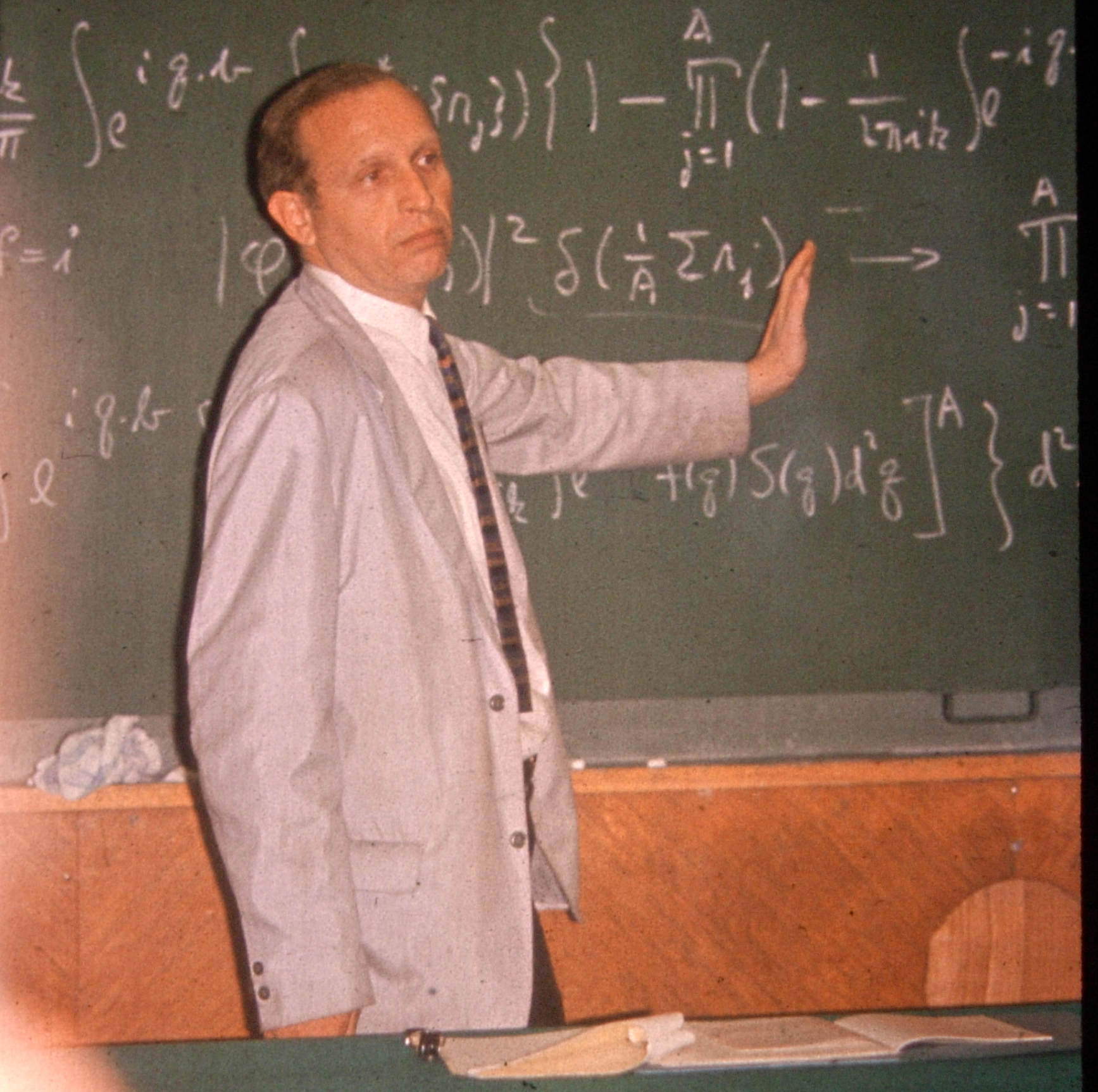}
\end{center}
%\vspace*{-10mm}
\caption{
Roy Glauber lecturing at a summer school in Hercegnovi, September 1969. Photo: P. Osland}
\label{Fig:hercegnovi}
\end{figure}

%%%%%%%%%%%%%%%%%%%%%%%%%%%%%%%%%%%%%%%%%%%%%%%%%%
I first met Roy Glauber at a Summer School in Hercegnovi (then Yugoslavia), in 1969. He was lecturing on scattering theory. This was of great interest to me, as I had done some multiple-scattering calculations for my thesis work in Trondheim. The formulas on the board (see Fig.~\ref{Fig:hercegnovi}) are reminiscent of his famous Boulder lectures \cite{Boulder}.

%%%%%%%%%%%%%%%%%%%%%%%%%%%%%%%%%%%%%%%%%%%%%%%%%%
%%%%%%%%%%%%%%%%%%%%%%%%%%%%%%%%%%%%%%%%%%%%%%%%%%

%%%%%%%%%%%%%%%%%%%%%%%%%%%%%%%%%%%%%%%%%%%%%%%%%%
\begin{figure}
\begin{center}
\includegraphics[scale=0.60]{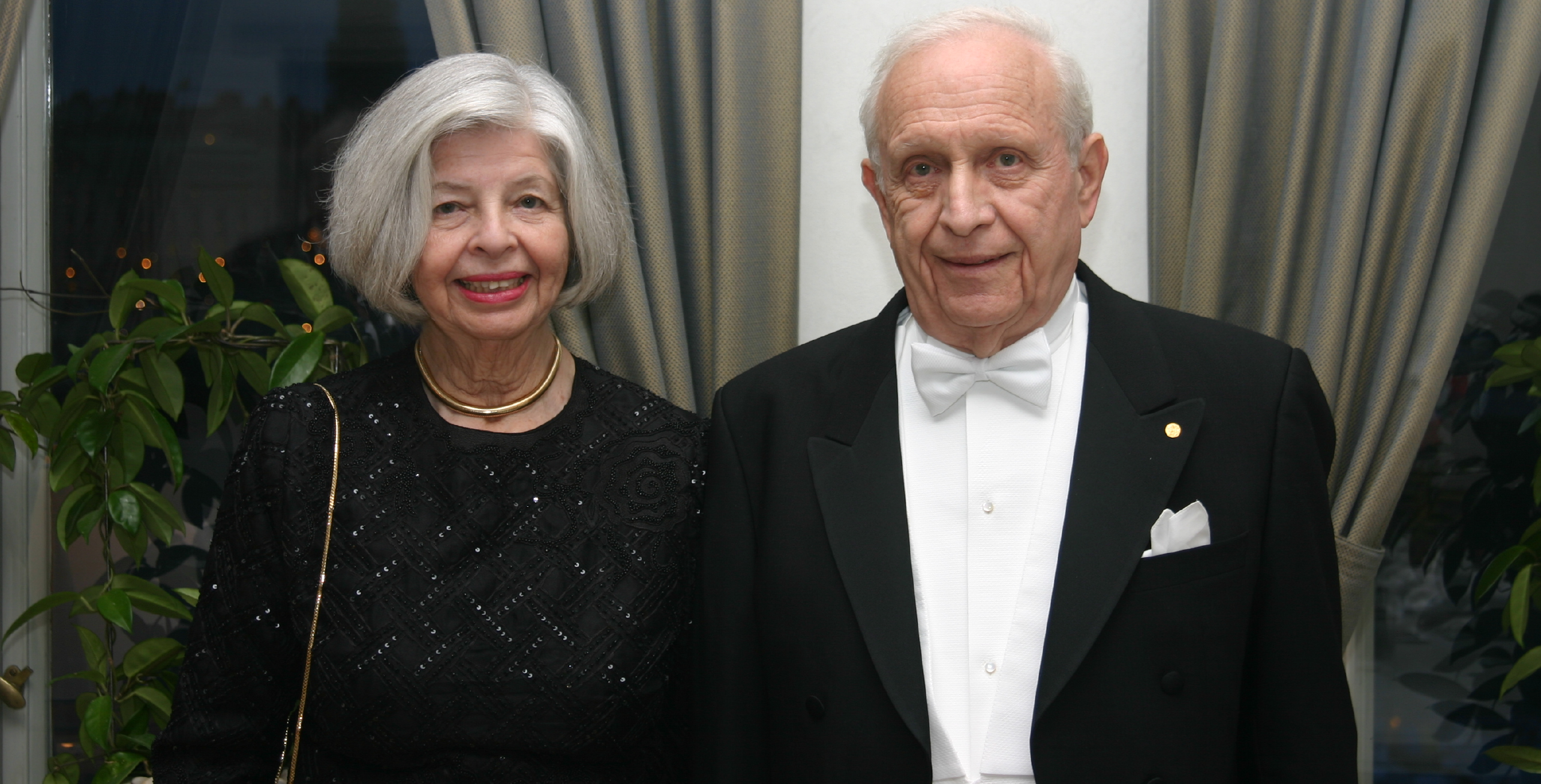}
\end{center}
%\vspace*{-10mm}
\caption{
Stockholm, December 2005. Photo: P. Osland}
\label{Fig:nobel}
\end{figure}
%%%%%%%%%%%%%%%%%%%%%%%%%%%%%%%%%%%%%%%%%%%%%%%%%%

%%%%%%%%%%%%%%%%%%%%%%%%%%%%%%%%%%%%%%%%%%%%%%%%%%
\begin{figure}
\begin{center}
\includegraphics[scale=0.23]{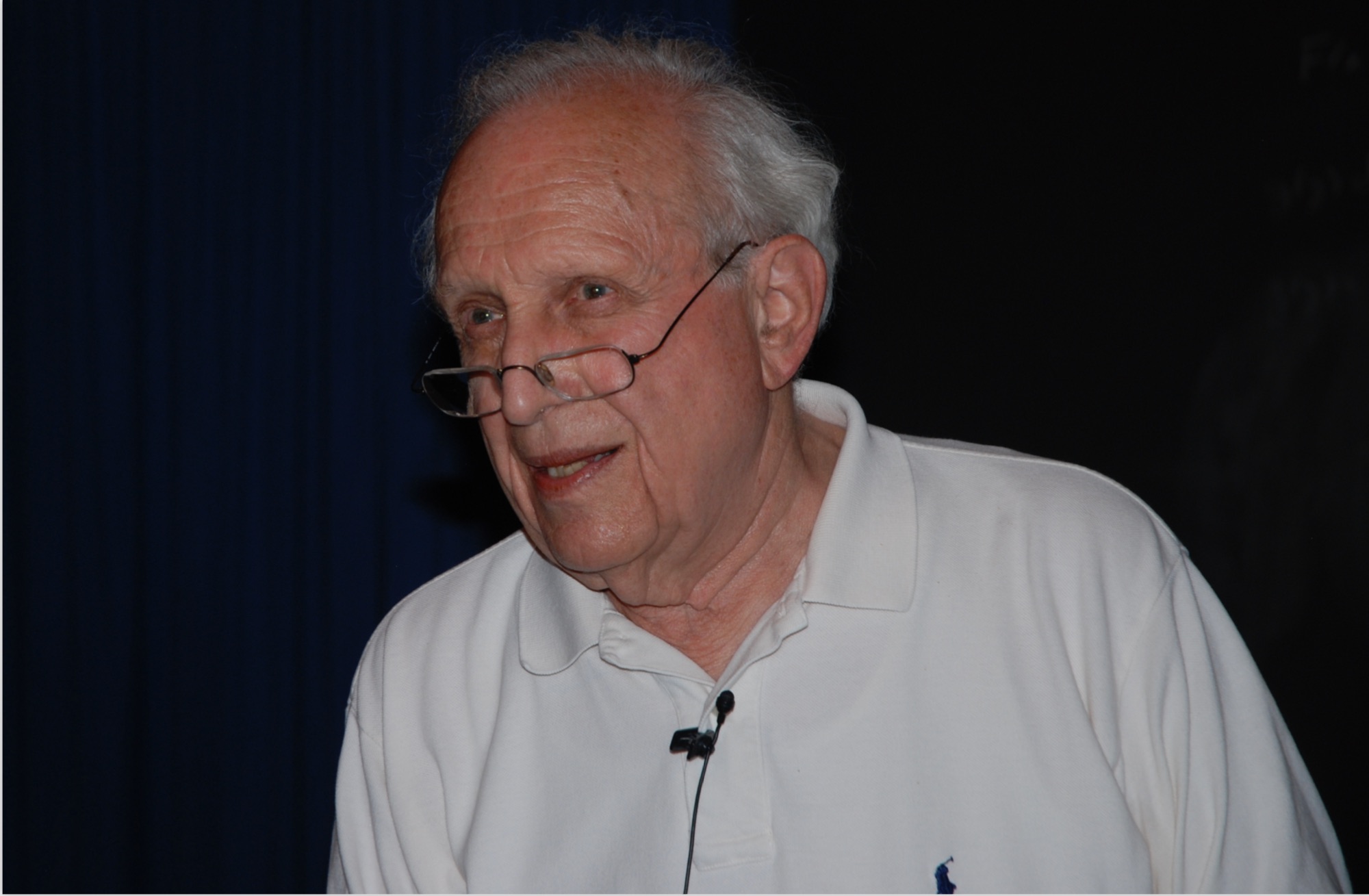}
\end{center}
%\vspace*{-10mm}
\caption{
CERN, Colloquium, August 2009. Photo: P. Osland}
\label{Fig:cern}
\end{figure}
%%%%%%%%%%%%%%%%%%%%%%%%%%%%%%%%%%%%%%%%%%%%%%%%%%

%%%%%%%%%%%%%%%%%%%%%%%%%%%%%%%%%%%%%%%%%%%%%%%%%%
\begin{figure}
\begin{center}
\includegraphics[scale=0.6]{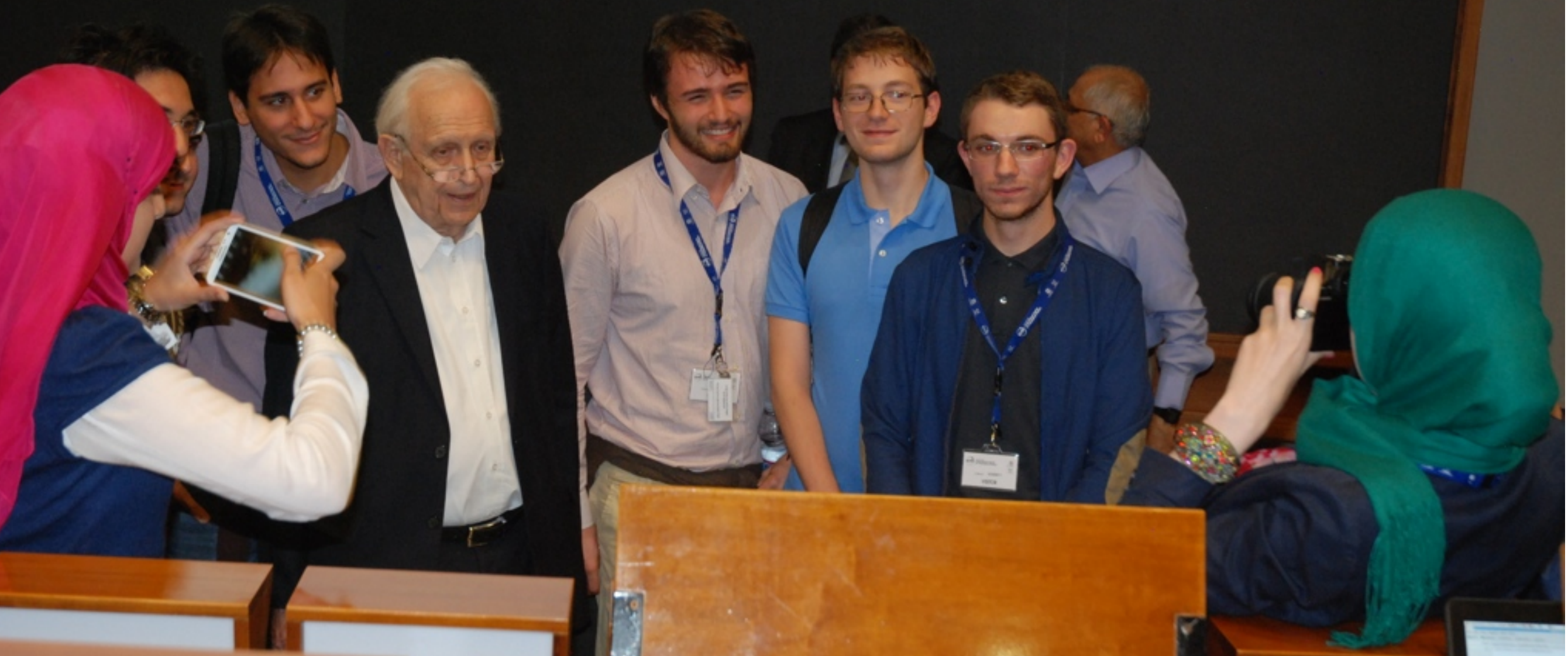}
\end{center}
%\vspace*{-10mm}
\caption{
Roy Glauber admired and photographed by the younger generation, at the ICTP 50-year celebration, October 2014. Photo: P. Osland}
\label{Fig:ictp}
\end{figure}
%%%%%%%%%%%%%%%%%%%%%%%%%%%%%%%%%%%%%%%%%%%%%%%%%%

%%%%%%%%%%%%%%%%%%%%%%%%%%%%%%%%%%%%%%%%%%%%%%%%%%
\begin{figure}
\begin{center}
\includegraphics[scale=0.25]{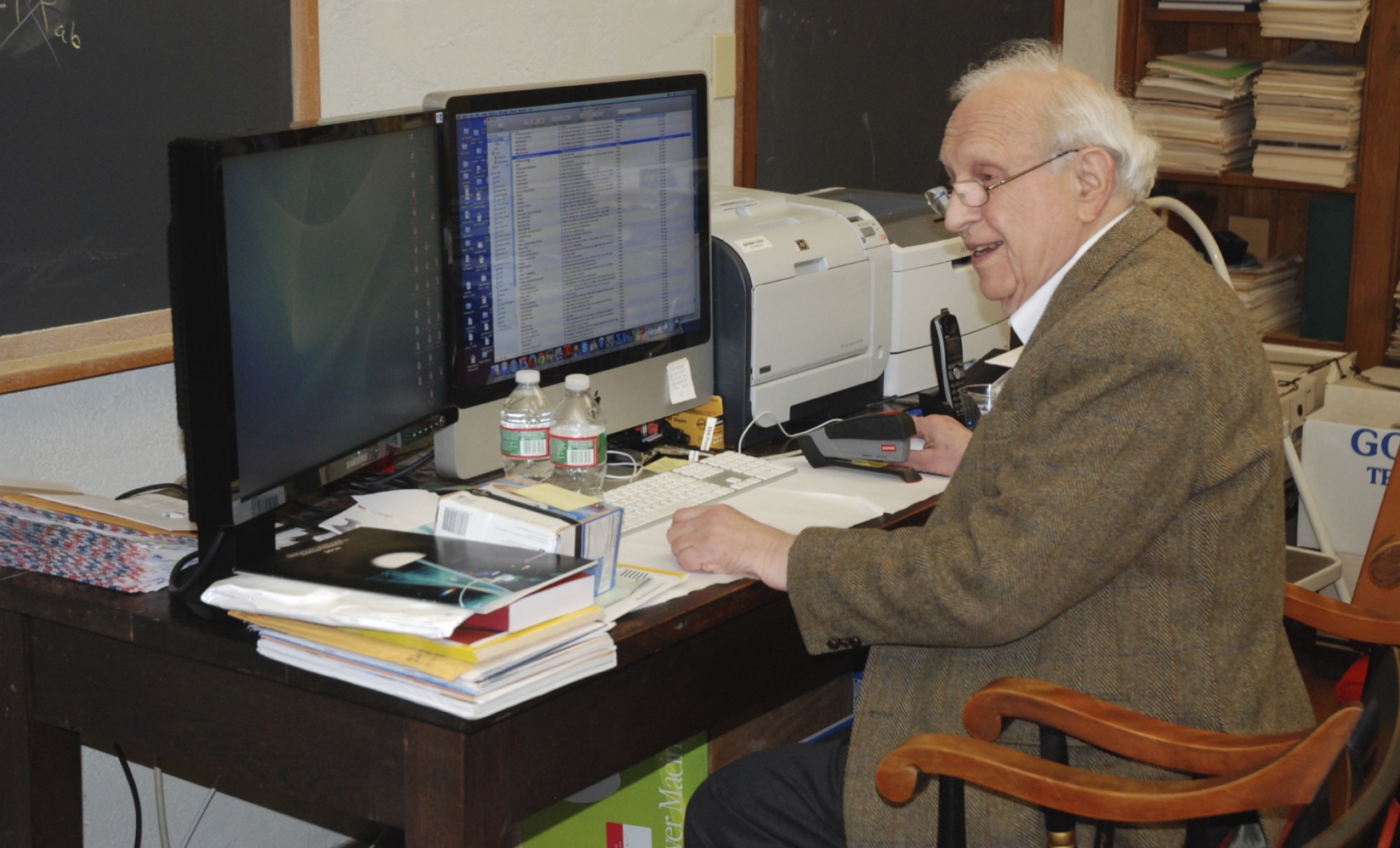}
\end{center}
%\vspace*{-10mm}
\caption{
Roy Glauber in his Lyman Lab office 331, January 2015. Photo: P. Osland}
\label{Fig:lyman}
\end{figure}
%%%%%%%%%%%%%%%%%%%%%%%%%%%%%%%%%%%%%%%%%%%%%%%%%%

I next had a year of overlap with him at CERN, around 1972, before being invited to Harvard, as a postdoc, in 1976, and then again in 1981. As mentioned at the beginning of Section~\ref{sect:diffraction}, this is when we started working on the asymptotic scattering, with Marek Bleszynski.
In the early 1990s Roy briefly visited me in Bergen two summers, continuing our work on ``the manuscript''.
But then we both got distracted by other ``urgent'' matters, and this work became another ``stack of papers'' in our offices.

In the aftermath of the Nobel Prize celebrations, in 2006, Roy suggested we should pick it up again. By then, the technology had thoroughly changed, all illustrations had to be provided in electronic form. This was a welcome opportunity to re-learn everything, and we continued our work during visits to Bergen, Harvard, and various European research centers, like CERN, Nordita and ICTP.

Roy had a fantastic memory. He had met all the significant physicists of his epoch, and had detailed and interesting stories to tell about them all. There exist recordings of him talking about the Manhattan Project \cite{Manhattan1}.

\end{appendix}

%\clearpage

% TODO:
% Provide your bibliography here. You have two options:

% FIRST OPTION - write your entries here directly, following the example below, including Author(s), Title, Journal Ref. with year in parentheses at the end, followed by the DOI number.
%\begin{thebibliography}{99}
%\bibitem{1931_Bethe_ZP_71} H. A. Bethe, {\it Zur Theorie der Metalle. i. Eigenwerte und Eigenfunktionen der linearen Atomkette}, Zeit. f{\"u}r Phys. {\bf 71}, 205 (1931), \doi{10.1007\%2FBF01341708}.
%\bibitem{arXiv:1108.2700} P. Ginsparg, {\it It was twenty years ago today... }, \url{http://arxiv.org/abs/1108.2700}.
%\end{thebibliography}

% \bibliographystyle{SciPost_bibstyle} % Include this style file here only if you are not using our template
\bibliography{SciPost_BiBTeX_File.bib}

\nolinenumbers

\end{document}